\documentclass[lettersize,journal,compsoc]{IEEEtran}
\usepackage[normalem]{ulem}
\usepackage{threeparttable}
\usepackage[stable]{footmisc}
\usepackage{makecell}
\usepackage{multirow}
\usepackage{bm}
\usepackage[justification=centering]{caption}
\usepackage{balance}
\usepackage{float}
\usepackage{tabularx}
\usepackage{longtable}
\usepackage{hyperref}
\usepackage{graphicx} 
\usepackage{xcolor,colortbl}%
\usepackage{orcidlink}
\usepackage{indentfirst} 
\usepackage{listings}
\usepackage{afterpage}
\usepackage{verbatim}
\usepackage{multirow}
\usepackage{array}
\usepackage{url}
\usepackage{booktabs}
\usepackage{pifont}
\usepackage{bbding}
\newcolumntype{L}{>{\centering\arraybackslash}m{5.2cm}}
\newcolumntype{R}{>{\centering\arraybackslash}m{2cm}}
\newcolumntype{A}{>{\centering\arraybackslash}m{3.5cm}}
\newcolumntype{B}{>{\centering\arraybackslash}m{1.7cm}}

\emergencystretch=\maxdimen
\definecolor{electricpurple}{rgb}{0.75, 0.0, 1.0}
\definecolor{lightgreen}{rgb}{0.90, 1.0, 0.90}
\definecolor{lightgray}{gray}{0.95}

\usepackage{ulem} 

\lstdefinelanguage{Kotlin}{
  morekeywords={val, var, fun, if, else, while, for, in, return, open, class, @Nullable},
  sensitive=true,
  morecomment=[l]{//},
  morecomment=[s]{/*}{*/},
  morestring=[b]",
  morestring=[s]{"""*}{*"""},
}

\lstdefinestyle{bw-diff}{
    basicstyle=\ttfamily\footnotesize, 
    numbers=none,
    tabsize=6,
    breaklines=false,                  
    columns=fullflexible,
    keepspaces=true,
    escapeinside={(*}{*)},
    captionpos=b,
    numberstyle=\color{safeSkyBlue}
}

\newcommand{\toolName}{\texttt{InteropScan}}
\newcommand{\toolDepend}{\texttt{DependExtractor}}
\newcommand{\projCount}{40}

\newcommand{\dquote}[1]{``#1"}

\begin{document}

\title{An Empirical Study of Kotlin-Java Cross-Dependency Issues and Their Detection}

\author{Qiong Feng~\orcidlink{0000-0003-1667-8062}, Xiaotian Ma~\orcidlink{0009-0008-9121-3656}, Jiayi Sheng~\orcidlink{0009-0009-4588-8780}, Huan Ji~\orcidlink{0009-0006-5636-9239}, Peng~Liang~\orcidlink{0000-0002-2056-5346}

\IEEEcompsocitemizethanks{

\IEEEcompsocthanksitem Qiong Feng, Xiaotian Ma, Jiayi Sheng are with Nanjing University of Science and Technology, China. Email:~\{qiongfeng, xyzboom,shengjiayi\}@njust.edu.cn

\IEEEcompsocthanksitem Huan Ji is with Huawei Technologies Co., Ltd., Shenzhen, Guangdong, China. Email:~jihuan3@huawei.com

\IEEEcompsocthanksitem Peng Liang is with School of Computer Science, Wuhan University, China. Email:~liangp@whu.edu.cn}

\thanks{Manuscript received; revised. (Corresponding author: Peng Liang)}}

\maketitle
\begin{abstract}

\noindent Since Google introduced Kotlin as an official programming language for developing Android apps in 2017, Kotlin has gained widespread adoption in Android development. The interoperability of Java and Kotlin's design nature allows them to coexist and interact with each other smoothly within a project. However, there is limited research on how Java and Kotlin interact with each other in real-world projects and what challenges are faced during these interactions. The answers to these questions are key to understanding these kinds of cross-language software systems. In this paper, we implemented a tool; named \toolDepend{}, which can extract 11 types and 9 types  of Kotlin-Java and Java-Kotlin dependencies, and conducted an empirical study of 40 Kotlin-Java real-world projects with 27,427 Java and 36,141 Kotlin source files. Our findings revealed that Java and Kotlin frequently interact with each other in these cross-language projects, with \texttt{Access} and \texttt{Call} dependency types being the most dominant. Compared to files interacting with other files in the same language, Java/Kotlin source files, which participate in the cross-language interactions, experience more commits and exhibit a higher defect rate. Additionally, among all Kotlin-Java problematic interactions, we identified ten common issues, along with their fixing strategies. Furthermore, we implemented a tool called \toolName{} to proactively detect these issues from Java and Kotlin source code. The findings of this study can help developers understand and address the challenges in Kotlin-Java projects.

\end{abstract}



\section{Introduction}
\label{sec:intro}

Since Google introduced Kotlin as an official programming language for developing Android apps in 2017, Kotlin has gained widespread adoption in Android development. According to recent studies, a significant of Android apps have been continuously migrated from Java to Kotlin~\cite{ardito:2020ist, mateus:2020esem}. The interoperability of Java and Kotlin's design nature allows them to coexist and evolve smoothly in a single project. Meanwhile, understanding the dependencies among software entities is fundamental to various aspects of software analysis, such as architecture recovery, code smell detection, and software quality evaluation~\cite{mo:2019tse,cui:2019icse,xiao:2021tse,feng:2019ase,feng:2023jsep}. Extracting and understanding the dependency interactions between Java and Kotlin code is essential for conducting software analysis in such cross-language software systems. Despite the increasing popularity of Kotlin and its interoperability with Java, there is limited research on how Java and Kotlin interact with each other in real-world projects. 


Furthermore, recent studies~\cite{li:2023understanding} have shown that a large percentage of bug fixes in deep learning frameworks implemented in Python and C++ involve modifying source code in both programming languages. Moreover, the complexity of code changes for such bug fixes is significantly higher than that of fixes in single-programming-language scenarios~\cite{li2023understanding}. Unlike the interactions between Python and C++, which rely on their compiled public APIs, Java and Kotlin, both being based on the Java Virtual Machine (JVM), can extensively depend on each other at the source code level. This raises the question of whether Kotlin-Java software systems also incur higher maintenance costs and are prone to cross-language bugs, similar to other multi-language systems like Python-C++. Moreover, objects passing through the Python-C interface can incur vulnerabilities due to the different rules in these two languages. For example, if handled incorrectly, a Python list passed to a C program can incur Buffer Overflow due to the fixed-size array in C~\cite{li:2022polycruise}. We are curious about the challenges developers face in the implementation and maintenance of Kotlin-Java systems and whether similar bugs or vulnerabilities exist in the Kotlin-Java interface. Understanding these challenges can assist researchers and practitioners in designing solutions and tools to navigate these issues and ensure the development of high-quality Kotlin-Java systems.

Although Google Android have provided documentation to guide developers on integrating Kotlin and Java~\cite{android-kotlin-java-interop}, the challenges faced in real-world projects remain largely unknown. In this paper, we propose to investigate into the dependency relations between Kotlin and Java in real-world projects to better understand the challenges associated with their interactions. To this end, we conducted an empirical study of \projCount{} open-source projects, comprising 27,427 Java and 36,141 Kotlin source files. We developed a static code analysis tool to extract 11 kinds of dependency relations between Kotlin and Java code entities within these projects. Our findings indicate that Java and Kotlin frequently interact in these cross-language projects, with \dquote{\texttt{Call}} and \dquote{\texttt{Access}} dependency types being the most prevalent (among 40 projects, 20.6\% to 30.0\% of cross-language dependencies are \dquote{\texttt{Call}}, and 49.6\% to 54.1\% of cross-language dependencies are \dquote{\texttt{Access}}). 

Compared to files interacting solely within the same language, Java/Kotlin source files that act as the interface for cross-language interactions have undergone more commits, but not necessarily contain a higher volume of lines of code (LOC). Furthermore, we identified nine common issues in Kotlin-Java problematic interactions, which deserve special attention from developers. To the best of our knowledge, this work is the first to explore Kotlin-Java dependency relations and to expose common problems in these interactions within real-world projects. The dependency analysis tool that we have implemented and the recurring patterns in Kotlin-Java problematic interactions can assist developers in comprehending and uncovering potential pitfalls in such cross-language systems. 

This paper extends our previous work~\cite{feng2024cross} by introducing a new research question (RQ4) and enhancing the results of RQ1, RQ2, and RQ3 through an increased volume and diversity of data, as well as a mixed-methods approach. Specifically, we analyzed 17 additional Kotlin-Java cross-language projects that are open-sourced on GitHub, exhibiting a varied ratio between Java and Kotlin code. With this expanded dataset, we validated our earlier findings and extended the previously identified seven categories of common Kotlin-Java interaction issues to nine categories. In addition, we improved our previous method for identifying relevant commits, moving from simple keyword searching to code change pattern matching, which significantly increased the precision of locating commits that address Kotlin-Java interaction issues. Furthermore, we developed a static code detection tool called \toolName{}, which leverages Kotlin compiler parsing features to proactively detect these common issues.

 
To summarize, our \textbf{contributions} in this paper are as follows:
\begin{itemize}
\item We implemented a static code analysis tool named \toolDepend{} to extract dependencies between Java and Kotlin, addressing a gap that existing static code analysis tools cannot resolve regarding such dependencies.
\item We demonstrated that the source files involved in Kotlin-Java interactions require undergoing more commits than those only involved in a single language. It also proves that the Kotlin-Java interface is actively changing.
\item We identified nine common issues which cause bugs and maintenance effort in Kotlin-Java problematic interactions. These common issues can serve as a taxonomy and deserve developers' special attention. 
\item We implemented a tool called \toolName{} to proactively detect Kotlin-Java interaction issues, achieving an 80.0\% detection rate.
\end{itemize}

\textbf{Data Availability Statement}: The data and implemented tool can be accessed through this link \cite{zenodo}.

\section{Approach}
\label{sec:approach}

\subsection{Kotlin-Java Dependency Extraction}
\label{approach:extraction}

\textbf{\toolDepend{} Tool:} We implemented a tool called \toolDepend{} in pure Kotlin to extract dependencies between Kotlin and Java code entities. \toolDepend{} first utilized the Kotlin compiler to resolve Kotlin and Java source files into PSI (Program Structure Interface, similar to Abstract Syntax Tree) elements, including their types, reference types, and locations. Subsequently, it extracted dependency relations between pairs of PSI elements. Finally, these dependencies  containing all PSI elements are aggregated into a graph and can be exported in JSON format, with each node representing a Kotlin or Java source file and each edge representing the dependencies between two source files. Following the definitions of dependency relations from related works~\cite{jin2019enre,depends,understand}, \toolDepend{} is capable of extracting 9 types of Java-Java dependencies, 14 types of Kotlin-Kotlin dependencies, 11 types of Kotlin-Java dependencies, and 9 types of Java-Kotlin dependencies. We list in Table~\ref{tbl:dp_types} all types of dependencies along with their descriptions (note that two types - Property Typed and Extension - occur only from Kotlin entities to Java or Kotlin entities, and three types - \texttt{Destructure Call}, \texttt{Class Delegate}, and \texttt{Property Delegate} - occur only from Kotlin entities to Kotlin entities). 

As observed from Table~\ref{tbl:dp_types}, \texttt{Access} signifies that an expression in one source file accesses a field or property of a class in another source file, while \texttt{Call} indicates that an expression in one source file calls a method of a class in another source file. Note that we not only record the dependency type between two source files, but also resolve dependency relations at a fine-grained granularity, capturing details such as the expression statement, locations, and type information for each specific dependency. This fine-grained approach can be valuable for future refactoring research, including but not limited to null safety addressing, exception handling, etc. Currently, \toolDepend{} supports Java version 17 and Kotlin version 1.9.22. For newer versions of Java and Kotlin, the tool only retains compatible functionality - newly introduced language features are not supported, though it remains capable of analyzing other source code that does not incorporate new features.

\begin{table}[!tbp]
	\centering
        \scriptsize
        \caption{Supported Dependency Types }
	\label{tbl:dp_types}
\begin{tabular}{|B|A|B|}
\hline
\textbf{DepType} & \textbf{Description} & Valid in                           \\ \hline  
\texttt{Access}     & an expression accesses a Kotlin/Java field or Kotlin property                         &  All \\ \hline
\texttt{Call}     & an expression calls a method &  All       \\ \hline
\texttt{Create}  & an expression creates an object &  All                   \\ \hline
\texttt{Extend}   & a class extends a parent class &  All                               \\ \hline
\texttt{Extension} & a Kotlin method or property extension & K-K, K-J                              \\ \hline
\texttt{Local Variable Typed (LVT)}   & a local variable uses a class as its type & All                                  \\ \hline
\texttt{Property Typed (PT)}      & a Kotlin property uses a Java class as its type & K-K, K-J                \\ \hline
\texttt{Implement}  & a class implements an interface & All            \\ \hline
\texttt{Import}    & a statement imports a type & All                                  \\ \hline
\texttt{Parameter}   & a method uses a class as its parameter &  All       \\ \hline
\texttt{Return}     & a method returns a class type &  All      \\ \hline
\texttt{Destructure Call (DC)} & a Kotlin destructure call. (e.g. \texttt{val (x, y) = obj}) & K-K \\ \hline
\texttt{Class Delegate (CD)} & a Kotlin class uses a Kotlin object as its delegate. & K-K \\ \hline
\texttt{Property Delegate (PD)} & a Kotlin property uses a Kotlin object as its delegate. & K-K \\ \hline
\end{tabular}
\end{table}
\textbf{\toolDepend{} Accuracy Validation:} Besides a large amount of unit testing, we also used a Kotlin-Java project with verified dependencies as the ground truth to conduct an integrated test. This ground truth project (containing 108 Java and 22 Kotlin source files) is our pre-release version of \toolDepend{}, implemented by using Java 1.8 and Kotlin 1.9.22 (\url{https://github.com/XYZboom/depends-kotlin/releases/tag/1.0.0-alpha0}). Just like software systems that migrate from Java to Kotlin, our tool also went through this process. As we are already familiar with the Kotlin and Java source files in this project, we feel confident in inspecting its dependencies. In addition, this project contains 5,424 dependency instances, including all 11 dependency types in K-J or J-K, making it a good candidate to test \toolDepend{}'s ability to extract various dependency instances. Specifically, two authors, each with 4 years of Java and 3 years of Kotlin development experience, traced entities line by line in each source file to list the dependency instances independently. After completing this, they discussed the results until an agreement was reached. This manual inspection served as ground truth and was compared with the JSON result generated by \toolDepend{}. The comparison shows that our tool can correctly capture 5,387 out of 5,424 dependencies (98.02\% precision and 99.34\% recall), increasing our confidence in applying \toolDepend{} in future experiments. Due to space limitations, the inspected ground truth and the results from the \toolDepend{} tool are shared in the provided Zenodo link~\cite{zenodo}.

\subsection{Common Issue Identification in Kotlin-Java Interaction}
\label{subsec:approach2}

To identify Kotlin-Java problematic interactions, we first narrowed the problematic commit scope and then conducted a thorough manual inspection of code changes to identify Kotlin-Java problematic patterns.

\textbf{Locate Commits to Address Kotlin-Java Problems}: 
The empirical study in our previous work~\cite{feng2024cross} shows that developers tend to apply specific code changes to resolve Kotlin-Java interoperability issues, such as adding \textit{Nullable} annotations. To precisely identify commits addressing these problems, we leverage \texttt{gumtree}~\cite{gumtree3}, a widely-used tool for analyzing code differences between two snapshots\cite{tufano2019empirical,jiang2018shaping,tsantalis2020refactoringminer}. \texttt{Gumtree} constructs abstract syntax trees (ASTs) for code in both pre- and post-change snapshots and detects changes to code entities, including \textit{insert}, \textit{delete}, \textit{update}, and \textit{move} operations. We use \texttt{gumtree} to extract code changes. If a commit contains any of these changes listed in Table~\ref{tbl:gumtree}, it is flagged for manual inspection.
\begin{table}[!tbp]
	\centering
        \scriptsize
        \caption{Extracted Commits with Code Change Types Signaling Problematic Kotlin-Java Interactions}
	\label{tbl:gumtree}
\begin{threeparttable}
\begin{tabular}{ccc}
\toprule
\textbf{Code Change Type} & \textbf{Code Details} & \textbf{Example Commit}~\tnote{a}      \\ 
\midrule  
Add Java Annotation      & \makecell{@Nullable \\ @NonNull} & \makecell{
\href{https://github.com/signalapp/Signal-Android/commit/6c56ef470f0fa6889335675f1a52a240796ec9c4}{Signal-Android 6ec9c4} \\ \hline
\href{https://github.com/androidx/androidx/commit/50ad9fab2058acc72add5484d2f80b1c7bf2c6b6}{androidx f2c6b6}
}\\ \hline
Add If Expression & if (xxx != null) & \href{https://github.com/signalapp/Signal-Android/commit/35c74573e77997401523ac8134f78f566c9f1ced}{Signal-Android 9f1ced} \\ \hline
Add Null Assertion & xxx!!           & \href{https://github.com/signalapp/Signal-Android/commit/fba9b46fe95d71df43acf5320f0d0192769b30d6}{Signal-Android 9b30d6} \\ \hline
Add Null Safe Operator & xxx?.xxx        &  \href{https://github.com/MatsuriDayo/NekoBoxForAndroid/commit/ed5b75e16cba769b9305eb55b084fbe362371cc8}{Neko...Android 371cc8} \\ \hline
Add Throws Annotation  & @Throws & \href{https://github.com/signalapp/Signal-Android/commit/e5d196c64254b0cd55fa326d11a8e6ece684ab09}{Signal-Android 84ab09} \\ \hline
\makecell{Change Non-Null \\ to Nullable} & \makecell{`Type' changed \\ to `Type?'} & \makecell{\href{https://github.com/nightscout/AndroidAPS/commit/5e28d9465db1d1cb904115cfa08d514ed41c1303}{AndroidAPS 1c1303}\\ 
\href{https://github.com/liangjingkanji/BRV/commit/2d7013903cca0e0708b9c98e1e5631304508f149}{BRV 08f149} }\\ \hline
\makecell{Change Mutability \\ of Set Types} & \makecell{Add or Remove \\ `Mutable' in Set Types} & \makecell{\href{https://github.com/AlmasB/FXGL/commit/d50d3f1fe46021486a170b4e50c2a5cc252860f9}{FXGL 2860f9} \\ \href{https://github.com/facebook/litho/commit/2062a2879568a4bbd37e54452c4a83953a36217e}{litho 36217e}} \\ 
\bottomrule
\end{tabular}
\begin{tablenotes}
      \item[a] Here are examples matching the code change types. However, whether they represent problematic Kotlin-Java interactions still requires manual validation.
\end{tablenotes}
\end{threeparttable}
\end{table}

\textbf{Identify Common Issues of Kotlin-Java Problems through Manual Inspection}: We establish a protocol for this manual inspection. First, we located the code changes in Kotlin or Java source file. Then we checked whether the code changes are related with Kotlin-Java interactions. We did this by tracing the code changes down to its reference type. For instance, if a line of code in a Kotlin file is deleted, we get the variables and methods in this line and track the variables down to check whether it is a type of Java class, method, etc. Subsequently, we check whether any new code snippets are added in the Kotlin file near the location of the deleted code line. If present, we compared the differences between the added line and the deleted line. Last, we check the commit comments and discussed what problems developers are trying to address and why developers made such changes. 

This step required the dedicated efforts of 2 students with 4 years of Java and 3 years of Kotlin development experience, and one senior researcher with 14 years of Java and 3 years of Kotlin development experience. We followed the practice described in \cite{campbell:2013coding}, allowing the senior researcher to establish the procedure for identifying code changes and training the students with examples. Then, the two students collaborated to identify code changes and categorize the changes' reasons. If disagreements occurred, the senior researcher would review the specific cases and discuss them with the two students until a consensus was reached.

\subsection{Automatic Detection Kotlin-Java Issue}

Based on the categorized Kotlin-Java common issues, we leveraged \toolDepend{} to extract Kotlin-Java dependencies and code entities at the statement level, and then implemented \toolName{} to enforce detection rules and identify these issues.

The misalignment between the Java and Kotlin type systems is the primary cause of Kotlin-Java interaction issues. Kotlin’s type system distinguishes between nullable and non-nullable types at compile time, ensuring null-safety---i.e., preventing runtime errors caused by missing values~\cite{KotlinSpec}. In contrast, Java’s type system does not make this distinction. All Java types without nullability annotations (e.g., \texttt{@NonNull} or \texttt{@Nullable}) are treated as platform types in Kotlin, whose nullability is uncertain---they may be either nullable or non-nullable.

\toolName{} leverages type analysis to identify potentially problematic interactions. Based on identified common patterns in Kotlin-Java interactions, \toolName{} implements specific detection rules to proactively catch these issues. For instance, if it detects that Kotlin code accesses Java code lacking nullability annotations without using a Null-Safe Call or Null-Safe Access, such interactions are flagged as potentially problematic.

\section{Study Design}
\label{sec:experiment}

\subsection{Subjects}

The objective of our study is to investigate the interactions between Kotlin and Java code within real-world projects, and we need to mine open source projects with both Kotlin and Java code. To achieve this, we first applied GitHub API to search repositories with Kotlin code through \href{https://github.com/search?q=language\%3AKotlin+\&type=repositories}{the GitHub link}. We further filtered those Kotlin repositories according to the following criteria:

\begin{itemize}
    \item [1)]The repository must contain Java code alongside Kotlin, with Java code constituting more than 3\% of the total code; 
    \item [2)]The repository should have a star count of no less than 20;
    \item [3)]The repository must have been updated within the last 48 hours relative to our data collection time, which is 12:00 PM on March 15, 2025.
\end{itemize}

\begin{table}[!th]
    \centering
    \caption{Language Distribution (\%) in Study Projects by Byte Size}
    \label{tbl:subjects-new}
    \begin{tabular}{ccc}
        \toprule
        \textbf{ProjectCount} & \textbf{Java(\%)} & \textbf{Kotlin(\%)} \\
        \midrule
        13 & 4.9-9.9 & 76.0-94.9\\
        7 & 10.5-19.0 & 79.7-89.1 \\
        8 & 20.1-27.7 & 45.9-78.2 \\
        8 & 31.4-36.9 & 61.9-67.9 \\
        4 & 42.2-48.7 & 49.5-57.3 \\
        \bottomrule
    \end{tabular}
    
\end{table}

Since GitHub does not support searching repositories by multiple languages, we clicked and manually inspected each repository to assess the ratio of programming languages and to fulfill Criterion 1). We established a minimum threshold of 3\% for Java code to ensure its active participation in the projects' functionalities, not just some random Java code. Therefore, we chose a 3\% threshold for Java code, rather than 1\% or 2\%. Criteria 2) and 3) focus on selecting repositories that are actively maintained, which are essential for identifying relevant challenges. By applying these criteria, we stopped the filtering once we had retrieved repositories from different domains. We identified \projCount{} projects, and their demographic information is presented in Table~\ref{tbl:subjects-new}.

These \projCount{} projects span various domains, including Android app development, Minecraft games, and IntelliJ IDEA plugins. For example, there are 13 projects with 4.9\%–9.9\% Java code and 76.0\%–94.9\% Kotlin code and 4 projects with 42.2\%–48.7\% Java code and 49.7\%–57.3\% Kotlin code. This wide and diverse distribution of Java and Kotlin code allows us to gain deeper insights into Kotlin–Java interactions.


\subsection{Research Questions}
We investigated the following Research Questions (RQs) to gain insight into Kotlin-Java interactions and challenges associated with their interactions.


\textbf{RQ1}. \textit{How prevalent are dependencies between Kotlin and Java in Kotlin-Java projects?} Kotlin is designed to be fully interoperable with Java, allowing code entities to be seamlessly accessed across both languages. Ideally, dependencies between Kotlin and Java should exhibit the same characteristics as those within homogeneous Kotlin or Java codebases. The answer to this question can enhance our understanding of such cross-language systems.

\textbf{RQ2}. \textit{Do files involved in Kotlin-Java interactions cost more commits and lines of code (LOCs) than files without Kotlin-Java interactions?} Recent studies on multi-language systems, such as those involving Python and C++, have shown that, to address bugs occurring in the interface of two languages, developers need to modify code in the interface, resulting in a higher number of lines of code (LOCs)~\cite{li2023understanding}. This observation raises the question of whether Kotlin-Java software systems are also prone to such cross-language maintenance activities. The answer to this question could pinpoint maintenance \texttt{hotspots} within these Kotlin-Java systems.

\textbf{RQ3}. \textit{Are there common problematic issues in Kotlin-Java interactions? and how to avoid or fix these issues in the interactions?} Previous research proposed various techniques for detecting code smells or anti-patterns within a single-language codebase. More recent studies have found that bugs and security problems in multiple language systems~\cite{li:2023understanding}. For instance, a recent study has uncovered security issues arising from interactions between Java and C code~\cite{li:2022polycruise}. By investigating whether there exist problematic issues in Kotlin-Java interactions, we aim to evaluate the problems associated with Kotlin-Java interactions and pinpoint the scenarios that may require closer attention when handling such interactions. 

\textbf{RQ4}. \textit{Can \toolName{} efficiently detect the problematic issues in Kotlin-Java interactions?} Based on the findings of RQ3, we implemented a prototype tool of \toolName{} to perform static analysis on the source code. This RQ assesses the effectiveness of \toolName{} in proactively detecting these interaction issues.
\section{Results}
\label{sec:results}

\subsection{Dependencies in Kotlin-Java Projects}
\label{subsec:rq1}

We applied \toolDepend{} on the most recent versions of our \projCount{} selected projects, and the results are presented in Table~\ref{tbl:rq1}. All dependencies in this table are counted at the file level, with expression-level dependencies aggregated to the file level. The last column shows the ratio of Kotlin-Java and Java-Kotlin dependencies to the minimum of Java-Java and Kotlin-Kotlin dependencies. According to the Barrel principle, if a project contains only a few Java source files, the number of cross-language dependencies should not be disproportionately large. Therefore, the smaller number of source files determines the cross-language dependencies. However, if we use the sum as the denominator, the resulting ratio would almost always be a small number below 1, making it difficult to distinguish projects with genuinely high levels of cross-language interaction. We compared Kotlin-Java and Java-Kotlin dependencies with the minimum of J-J and K-K to check whether cross-language dependencies are more prevalent than single-language dependencies.

\begingroup{
\rowcolors{3}{gray!10}{white}
\begin{table}[!htbp]
\scriptsize
\centering
\caption{Overall Dependencies in the Subjects}
\scriptsize
\label{tbl:rq1}
\resizebox{\linewidth}{!}{
    \begin{tabular}{c|cccccc|c}
        \toprule
        \multirow{2}{*}{\textbf{Project}} & \multirow{2}{*}{\textbf{\#J}} & \multirow{2}{*}{\textbf{\#K}} & \multirow{2}{*}{\textbf{J-J}} & \multirow{2}{*}{\textbf{K-K}} & \multirow{2}{*}{\textbf{J-K}} & \multirow{2}{*}{\textbf{K-J}} & \multirow{2}{*}{\textbf{$\frac{K-J + J-K}{Min(J-J,K-K)}$\%}} \\
        & & & & & & & \\
        \midrule
        AndroidAPS & 9179 & 2979 & 10147 & 53288 & 93655 & 1177 & 934.6 \\
        Androidx & 6557 & 6599 & 100954 & 116191 & 2033 & 14935 & 16.8 \\
        Auxio & 3 & 282 & 17 & 3941 & 0 & 27 & 158.82 \\
        anchor-platform & 536 & 249 & 6292 & 819 & 5 & 4796 & 586.2 \\
        BaseRecyclerViewAdapterHelper & 23 & 61 & 58 & 435 & 82 & 113 & 336.2 \\
        BiliRoamingX & 209 & 384 & 536 & 2242 & 79 & 609 & 128.4 \\
        BRV & 4 & 114 & 0 & 525 & 27 & 11 & - \\
        bilimiao2 & 86 & 454 & 917 & 5983 & 0 & 121 & 13.2 \\
        bugsnag-android & 132 & 455 & 665 & 2862 & 1030 & 1592 & 394.3 \\
        DanDanPlayForAndroid & 52 & 583 & 286 & 6257 & 4 & 92 & 33.6 \\
        DataStats & 8 & 10 & 11 & 24 & 2 & 46 & 436.4 \\
        dgs-framework & 99 & 293 & 264 & 1648 & 117 & 570 & 260.2 \\
        EhViewer & 151 & 179 & 868 & 2157 & 82 & 915 & 114.9 \\
        FDPClient & 209 & 619 & 1017 & 8860 & 1594 & 1279 & 282.5 \\
        Flap & 16 & 93 & 58 & 638 & 0 & 19 & 32.8 \\
        FXGL & 418 & 464 & 3995 & 3610 & 2452 & 1445 & 107.9 \\
        facebook-android-sdk & 199 & 444 & 1270 & 5399 & 2389 & 360 & 216.5 \\
        grpc-kotlin & 6 & 77 & 6 & 285 & 11 & 8 & 316.7 \\
        habitica-android & 12 & 802 & 6 & 10866 & 152 & 179 & 5516.7 \\
        husi & 29 & 171 & 214 & 1931 & 45 & 494 & 251.9 \\
        intellij-powershell & 6 & 155 & 11 & 930 & 53 & 12 & 590.9 \\
        javalin & 54 & 260 & 174 & 2595 & 305 & 629 & 536.8 \\
        jetbrains & 86 & 291 & 283 & 2107 & 201 & 204 & 143.1 \\
        ka-testing & 7 & 70 & 12 & 170 & 33 & 3 & 300.0 \\
        legado & 84 & 722 & 833 & 12154 & 0 & 90 & 10.8 \\
        LiquidBounce & 78 & 565 & 30 & 9117 & 735 & 11 & 2486.7 \\
        litho & 1011 & 75 & 15639 & 182 & 0 & 1148 & 630.8 \\
        MyExpenses & 313 & 698 & 2091 & 9745 & 815 & 2841 & 174.8 \\
        maps & 171 & 39 & 830 & 340 & 603 & 191 & 233.5 \\
        muzei & 5 & 162 & 3 & 1340 & 0 & 9 & 300.0 \\
        NekoBoxForAndroid & 30 & 164 & 220 & 1899 & 28 & 519 & 248.6 \\
        NightX-Client & 171 & 251 & 845 & 3221 & 1014 & 1242 & 267.0 \\
        RootEncoder & 131 & 330 & 990 & 3209 & 589 & 393 & 99.2 \\
        SeriesGuide & 168 & 391 & 788 & 4508 & 575 & 1472 & 259.8 \\
        Signal-Android & 1990 & 2717 & 20180 & 36406 & 13252 & 16994 & 149.9 \\
        SkyHanni & 296 & 692 & 819 & 14258 & 630 & 1898 & 308.7 \\
        SkytilsMod & 87 & 299 & 19 & 3636 & 160 & 107 & 1405.3 \\
        simbrain & 290 & 468 & 2068 & 8341 & 2006 & 1443 & 166.8 \\
        tasks & 81 & 631 & 165 & 9029 & 184 & 1242 & 864.2 \\
        wire & 195 & 581 & 561 & 9931 & 4449 & 61 & 803.9 \\
        \midrule
        \rowcolor{white}
        \multicolumn{1}{c}{average}  & \multicolumn{6}{c}{} & \textbf{515.9} \\
        \bottomrule
    \end{tabular}
}

\caption*{\scriptsize \tiny \#J and \#K denote the number of Java and Kotlin files; J-J, K-K, K-J and J-K denote the dependencies number from Java to Java, Kotlin to Kotlin, Kotlin to Java and Java to Kotlin entities, respectively.}
    \vspace{-4mm}
\end{table}
}\endgroup

\begin{figure*}
    \centering
    \includegraphics[width=0.9\textwidth]{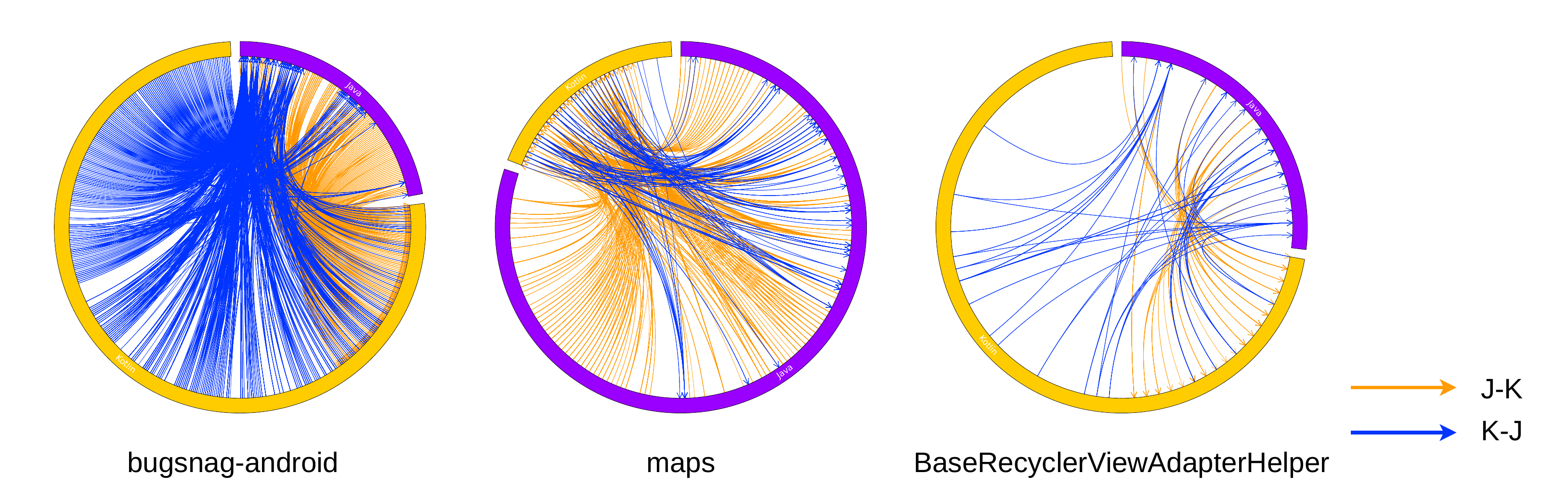}
    \caption{Circular Layout of Dependencies}
    \label{fig:deps}
\end{figure*}

As shown in Table~\ref{tbl:rq1}, Kotlin and Java source files frequently interact. We identified both Java-Kotlin and Kotlin-Java dependencies in 35 projects, and Kotlin-Java dependencies in 5 projects: {\sffamily bilimiao2}, {\sffamily Flap}, {\sffamily legado}, {\sffamily litho}, {\sffamily muzei}. For instance, in the {\sffamily bilimiao2} project, there are 121 Kotlin-Java dependencies within 86 Java and 454 Kotlin files. If we divide 121 Kotlin-Java dependencies by the minimum of Java-Java and Kotlin-Kotlin dependencies, we obtain 13.2\%. This significant ratio indicates that Kotlin-Java dependencies are prevalent in this project. Similar results are observed in other projects. In the {\sffamily DataStats} project, for example, there are 2 Java-Kotlin and 46 Kotlin-Java dependencies within 8 Java and 10 Kotlin files, outnumbering 11 Java-Java and 24 Kotlin-Kotlin dependencies. The resulting ratio of 436.4\% ((2+46)/11) underscores the dominance of Kotlin-Java dependencies in this project. Overall, the last column of this table reveals that the ratio of Kotlin-Java dependencies ranges from 10.8\% in the {\sffamily legado} project to 5516.7\% in the {\sffamily habitica-android} project, with an average of 515.9\%. This average indicates that, on average, Kotlin-Java interactions are five times more prevalent than the minority of single-language interactions. Also, Table~\ref{tbl:rq1} shows that there are more J-K dependencies than K-J in {\sffamily AndroidAPS}, {\sffamily FXGL}, {\sffamily RootEncoder} while there are less J-K dependencies than K-J in {\sffamily Androidx}, {\sffamily EhViewer}, {\sffamily Signal-Android}. We hypothesize that variations in J–K and K–J dependency patterns may reflect different migration strategies. We conducted a longitudinal analysis of six projects sampled from the subjects in Table~\ref{tbl:rq1}: three projects dominated by K-J dependencies (\textsf{EhViewer}, \textsf{BiliRoamingX}, \textsf{anchor-platform}) and three dominated by J-K dependencies (\textsf{wire}, \textsf{maps}, \textsf{LiquidBounce}). For each project we applied \texttt{DependExtractor} to approximately 20 tagged releases spanning the project's history (115 release snapshots in total). Our longitudinal analysis proves that the directional dominance ($J\rightarrow K$ vs. $K\rightarrow J$) is explicitly driven by a project's migration strategy. Projects utilizing a \textit{bottom-up strategy} (\textit{wire}, \textit{maps}) convert low-level foundational libraries to Kotlin first; because the business logic remains in legacy Java, this creates massive $J\rightarrow K$ dependencies. Conversely, projects adopting a \textit{top-down strategy} (\textit{anchor-platform}, \textit{EhViewer}) rewrite upper business layers in Kotlin first, forcing these new components to constantly call down into un-migrated Java cores, resulting in a high percentage of $K\rightarrow J$ dependencies.


Figure~\ref{fig:deps} depicts the circular layout of dependencies in three projects. The yellow and purple edges represent Kotlin and Java source files, respectively. The yellow and blue curved lines indicate Java-to-Kotlin and Kotlin-to-Java dependencies, respectively. Note that the curved lines shown in Figure~\ref{fig:deps} do not have weights, meaning that they illustrate only the existence of dependencies rather than the strength of the dependencies between Kotlin and Java source files. It is evident that in these three projects, Kotlin–Java dependencies are not confined to a specific area; instead, they are distributed throughout the entire system. We observed this pattern in 31 out of 40 projects. For the remaining 9 projects, we conducted a manual inspection and found that the narrower scope of Java–Kotlin and Kotlin–Java dependencies is typically due to one language serving as a utility or shared library in the migration process.

We calculated statistics of Kotlin-Java dependency types in each project and the results showed that certain dependency types are present in some projects but not in others. For instance, the ``\texttt{Extension}'' dependency type is found in 16 projects, while the remaining 24 projects do not exhibit this type. We suspect this discrepancy is due to the distinct characteristics of the projects. However, all 11 defined types of Kotlin-Java dependencies can be observed across all projects. The top two most frequent dependency types among Kotlin-Java interactions are ``\texttt{Access}'' and ``\texttt{Call}''.

\textbf{To answer RQ1, the dependency extraction from \projCount{} open-source Kotlin-Java projects spanning various domains reveals that Kotlin-Java interactions are widespread. Only 6 projects (Androidx, bilimiao2, DanDanPlayForAndroid, Flap, legado, RootEncoder) has cross-lang dependencies less than non-cross-lang dependencies. On average, the cross-lang dependencies are about 5 times of non-cross-lang dependencies among 40 projects. Thus validating the interoperability between Kotlin and Java.} Kotlin-Java and Java-Kotlin dependencies can occur more frequently than minor single-language interactions, with ``\texttt{Access}'' and ``\texttt{Call}''  being the top two most frequent dependency types. Furthermore, in most projects Kotlin-Java dependencies are distributed across the entire system rather than confined to a local area.

\definecolor{tblheader}{gray}{0.75} 
\definecolor{cmtgray}{gray}{0.82}   
\definecolor{Churngray}{gray}{0.92}   

\newcommand{\chg}{\cellcolor{cmtgray}} 
\newcommand{\lhg}{\cellcolor{Churngray}} 

\begin{table*}[!htbp]
    \centering
    \caption{Maintenance Cost of Kotlin/Java Source Files (not) in Cross-Language (CL) Dependencies}
    \label{tbl:rq2-maintenance-cost}
    \begin{tabular}{l cccccc ccc}
        \toprule
        \textbf{ProjectName} & \multicolumn{2}{c}{\textbf{Java not in CL}} & \multicolumn{2}{c}{\textbf{Java in CL}} & \multicolumn{2}{c}{\textbf{Kotlin not in CL}} & \multicolumn{2}{c}{\textbf{Kotlin in CL}} \\ 
        \cmidrule(lr){2-3} \cmidrule(lr){4-5} \cmidrule(lr){6-7} \cmidrule(lr){8-9}
        
        & \textbf{\#Commit} & \textbf{\#Churn} & \textbf{\#Commit} & \textbf{\#Churn} & \textbf{\#Commit} & \textbf{\#Churn} & \textbf{\#Commit} & \textbf{\#Churn} \\ 
        \midrule
        {\sffamily AndroidAPS} & 7.8 & 86.3 & 0.3 & 3.9 & 11.5 & 121.6 & \chg 21.5(1.9x) & \lhg 201.0 \\ 
        {\sffamily Androidx} & 7.1 & 251.6 & \chg 12.1(1.7x) & \lhg 509.1 & 11.7 & 218.6 & 10.0 & \lhg 240.6 \\ 
        {\sffamily Auxio} & - & - & 8.0 & 980.7 & 32.3 & 274.9 & \chg 60.7(1.9x) & \lhg 435.5 \\ 
        {\sffamily anchor-platform} & 3.9 & 22.8 & \chg 5.4(1.4x) & \lhg 37.5 & 4.6 & 26.1 & \chg 6.2(1.3x) & \lhg 96.3 \\ 
        {\sffamily BaseRecyclerViewAdapterHelper} & 5.8 & 198.3 & \chg 8.6(1.5x) & 143.4 & 6.4 & 87.5 & \chg 15.0(2.3x) & \lhg 205.3 \\ 
        {\sffamily BRV} & - & - & 6.8 & 630.3 & 7.1 & 81.3 & \chg 27.8(3.9x) & \lhg 169.2 \\ 
        \addlinespace[0.6em] 
        {\sffamily BiliRoamingX} & 2.2 & 54.8 & \chg 7.2(3.3x) & \lhg 163.4 & 2.4 & 50.8 & \chg 10.9(4.5x) & \lhg 164.2 \\ 
        {\sffamily bilimiao2} & 1.5 & 5.5 & 0.5 & 1.7 & 5.3 & 37.2 & \chg 16.1(3.0x) & \lhg 110.4 \\ 
        {\sffamily bugsnag-android} & 4.6 & 90.3 & \chg 28.3(6.1x) & \lhg 214.9 & 3.8 & 78.6 & \chg 9.4(2.5x) & \lhg 87.9 \\ 
        {\sffamily DanDanPlayForAndroid} & 1.9 & 167.7 & \chg 2.4(1.3x) & \lhg 283.3 & 4.4 & 94.0 & \chg 6.6(1.5x) & \lhg 144.6 \\ 
        {\sffamily DataStats} & 1.5 & 9.0 & \chg 4.7(3.1x) & \lhg 28.8 & - & - & 2.8 & 16.3 \\ 
        \addlinespace[0.6em]
        {\sffamily dgs-framework} & 1.8 & 11.0 & \chg 2.7(1.5x) & \lhg 16.2 & 4.5 & 29.2 & \chg 9.4(2.1x) & \lhg 82.7 \\ 
        {\sffamily EhViewer} & 2.4 & 134.4 & \chg 3.9(1.7x) & \lhg 185.1 & 11.2 & 153.9 & \chg 42.1(3.7x) & \lhg 499.5 \\ 
        {\sffamily FDPClient} & 4.6 & 19.2 & \chg 13.0(2.8x) & \lhg 91.9 & 4.6 & 17.9 & \chg 8.6(1.9x) & \lhg 46.1 \\ 
        {\sffamily Flap} & 7.7 & 39.1 & 3.3 & 8.3 & 8.3 & 48.4 & \chg 13.3(1.6x) & \lhg 102.1 \\ 
        {\sffamily FXGL} & 7.0 & 147.1 & \chg 9.5(1.3x) & \lhg 208.6 & 7.1 & 116.8 & \chg 11.4(1.6x) & \lhg 195.3 \\ 
        \addlinespace[0.6em]
        {\sffamily facebook-android-sdk} & 4.7 & 103.6 & \chg 10.3(2.2x) & \lhg 273.4 & 6.9 & 156.4 & 6.7 & \lhg 229.9 \\ 
        {\sffamily grpc-kotlin} & 2.0 & 5.0 & 1.0 & 2.8 & 3.0 & 24.5 & \chg 4.3(1.5x) & 20.0 \\ 
        {\sffamily habitica-android} & 39.0 & 4721.0 & 11.5 & 362.7 & 16.8 & 200.9 & \chg 28.5(1.7x) & \lhg 385.1 \\ 
        {\sffamily husi} & - & - & 2.9 & 28.3 & 3.1 & 16.9 & \chg 3.3(1.0x) & \lhg 19.7 \\ 
        {\sffamily intellij-powershell} & 11.5 & 141.0 & 7.3 & 54.0 & 3.7 & 20.5 & \chg 8.4(2.3x) & \lhg 39.5 \\ 
        \addlinespace[0.6em]
        {\sffamily javalin} & 4.0 & 11.0 & \chg 4.8(1.2x) & \lhg 25.8 & 6.6 & 30.8 & \chg 10.5(1.6x) & \lhg 57.5 \\ 
        {\sffamily jetbrains} & 4.1 & 37.1 & \chg 7.7(1.9x) & \lhg 49.6 & 4.1 & 21.0 & \chg 9.5(2.3x) & \lhg 54.1 \\ 
        {\sffamily karibu-testing} & - & - & 7.6 & 59.4 & 9.7 & 82.8 & \chg 30.3(3.1x) & \lhg 219.7 \\ 
        {\sffamily LiquidBounce} & 2.4 & 11.2 & \chg 9.5(4.0x) & \lhg 70.9 & 7.9 & 40.9 & \chg 13.2(1.7x) & \lhg 65.5 \\ 
        \addlinespace[0.6em]
        {\sffamily legado} & 2.4 & 8.1 & 1.3 & 6.5 & 26.5 & 139.5 & \chg 41.9(1.6x) & \lhg 825.3 \\ 
        {\sffamily litho} & 26.7 & 205.6 & \chg 74.2(2.8x) & \lhg 439.3 & 0.0 & 0.0 & 0.0 & 0.0 \\ 
        {\sffamily MyExpenses} & 2.2 & 76.5 & \chg 26.7(12.3x) & \lhg 140.6 & 5.7 & 33.1 & \chg 15.3(2.7x) & \lhg 86.3 \\ 
        {\sffamily maps} & 0.6 & 14.3 & 0.5 & \lhg 19.2 & 0.0 & 0.0 & 0.0 & 0.0 \\ 
        {\sffamily muzei} & 7.5 & 123.5 & 6.3 & \lhg 939.7 & 14.7 & 208.5 & \chg 40.3(2.7x) & \lhg 386.7 \\ 
        \addlinespace[0.6em]
        {\sffamily NightX-Client} & 3.4 & 15.2 & \chg 6.3(1.9x) & \lhg 43.0 & 4.4 & 20.9 & \chg 5.3(1.2x) & \lhg 28.2 \\ 
        {\sffamily NekoBoxForAndroid} & 0.0 & 0.0 & \chg 1.8(+) & \lhg 31.1(+) & 2.5 & 16.6 & \chg 5.8(2.4x) & \lhg 39.8 \\ 
        {\sffamily RootEncoder} & 4.3 & 23.3 & \chg 14.1(3.3x) & \lhg 85.6 & 4.7 & 31.1 & \chg 8.5(1.8x) & \lhg 50.1 \\ 
        {\sffamily SeriesGuide} & 2.2 & 9.8 & \chg 6.0(2.7x) & \lhg 39.4 & 4.4 & 25.2 & \chg 6.9(1.6x) & \lhg 43.7 \\ 
        {\sffamily Signal-Android} & 3.3 & 67.6 & \chg 14.8(4.5x) & \lhg 167.8 & 2.9 & 72.6 & \chg 8.0(2.8x) & \lhg 155.1 \\ 
        \addlinespace[0.6em]
        {\sffamily SkyHanni} & 3.0 & 16.8 & \chg 3.6(1.2x) & \lhg 23.1 & 7.1 & 39.1 & \chg 20.3(2.9x) & \lhg 107.9 \\ 
        {\sffamily SkytilsMod} & 6.5 & 24.0 & 2.9 & 20.1 & 5.6 & 22.9 & \chg 10.2(1.8x) & \lhg 54.5 \\ 
        {\sffamily simbrain} & 4.3 & 95.9 & \chg 7.7(1.8x) & 57.2 & 4.8 & 46.5 & \chg 9.0(1.9x) & \lhg 66.3 \\ 
        {\sffamily tasks} & 5.3 & 38.2 & \chg 9.2(1.7x) & \lhg 59.1 & 5.9 & 31.6 & \chg 10.4(1.7x) & \lhg 58.3 \\ 
        {\sffamily wire} & 1.0 & 8.0 & \chg 1.7(1.7x) & \lhg 13.6 & 5.4 & 38.0 & 3.9 & 16.3 \\ 
        \bottomrule
    \end{tabular}
    
    \smallskip
    \scriptsize
    \colorbox{cmtgray}{\quad} Darker shading indicates the \textbf{\#Commit} of Kotlin/Java source files in CL is larger than those not in CL; \\ 
    \colorbox{Churngray}{\quad} Lighter shading indicates the \textbf{\#Churn} of Kotlin/Java source files in CL is larger than those not in CL.
\end{table*}
\subsection{Maintenance Cost in Kotlin-Java Projects }
\label{subsec:rq2}

Recent studies on multi-language systems have revealed that the interface between two languages is vulnerable and susceptible to bugs, often resulting in increased maintenance costs for fixing these issues~\cite{li2023understanding,li:2022polycruise,li:2023understanding}. This raises the question of whether the interface between Kotlin and Java faces similar challenges. 

While the number of bugs could be an indicative metric for studying Kotlin–Java interaction problems, we found that some projects do not use a bug-labeling system. For example, in the Signal-Android project, there are no issues labeled as bugs. Therefore, we used commit counts and modified lines of code (churn) to examine whether files involved in Kotlin–Java interactions tend to be more maintainable. For each project, utilizing the dependency information obtained in RQ1 (see Section~\ref{subsec:rq1}), we extracted the paths of Kotlin or Java source files involved in Kotlin-Java or Java-Kotlin dependencies, denoted as \textit{Paths(kt/java)}. Subsequently, we traversed each project's git commits and matched each modified file path of a commit with \textit{Paths(kt/java)}. If a match was found, we incremented \textit{\#Commit(aFile)} by 1 and added the modified lines of code (including insertions and deletions) to \textit{\#Churn(aFile)}. Similarly, we calculated \textit{\#Commit} and \textit{\#Churn} for Java or Kotlin source files participating only in Java-Java or Kotlin-Kotlin interactions.

Table~\ref{tbl:rq2-maintenance-cost} presents the average churn and commits for different types of files in each project. The first two columns (Java not in CL) denote the average commit counts and churn for Java source files that do not interact with Kotlin source files. The second two columns (Java in CL) denote the average commit counts and churn for Java source files that interact with Kotlin source files, either in Java-Kotlin or Kotlin-Java dependencies. Similarly, the third two columns (Kotlin not in CL) and the last two columns (Kotlin in CL) denote Kotlin source files that does not and does interact with Java source files. 

Comparing the fourth-to-last column with the second-to-last column, we observe that in 34 out of 40 projects, the average number of commits for Kotlin source files participating in Kotlin-Java interactions is significantly larger than for Kotlin source files not participating in Kotlin-Java interactions. The number in the parentheses denotes the ratio of these two commit counts. For example, in the {\sffamily karibu-testing} project, the average number of commits (30.3) for Kotlin source files participating in Kotlin-Java interactions is 3.1 times of that (9.7) for Kotlin source files not participating in Kotlin-Java interactions. Similarly, in 26 out of 40 projects, the average number of commits for Java source files participating in Kotlin-Java interactions is significantly larger than for Java source files not participating in Kotlin-Java interactions.

As observed from Table~\ref{tbl:rq2-maintenance-cost}, the churn for Java source files participating in Kotlin-Java interactions show some difference from Java source files not participating in Kotlin-Java interactions. In 26 out of 40 projects and 35 out of 40 projects, Java and Kotlin source files participating in Kotlin-Java interactions have more churns on average, respectively. Furthermore, we conducted a t-test~\cite{rice2007mathematical} for these columns. As for the nonexistent values (- in the table), we skipped them when doing tests. The t-test results between column \#Commit (Kotlin in CL) and \#Commit (Kotlin not in CL) indicate a statistically significant difference at a 99\% confidence level (\textit{p-value} = 0.001, \textit{Cohen's d}=0.909). However, the t-test results indicate no statistically significant difference between column \#Commit (Java in CL) and \#Commit (Java not in CL) (\textit{p-value} = 0.064, \textit{Cohen's d}=0.323). Such results support the finding that the commits of Kotlin source files in cross-language is significantly greater than that of Kotlin not in Cross-Language while the commits of Java source files in or not in Cross-Language do not show a significant difference. Furthermore, the t-test results between column \#Churn (Kotlin in CL) and \#Churn (Kotlin not in CL) indicate a statistically median difference at a 99\% confidence level (\textit{p-value} = 0.001, \textit{Cohen's d}=0.599). However, the t-test results indicate no statistically significant difference between column \#Churn (Java in CL) and \#Churn (Java not in CL) (\textit{p-value} = 0.630, \textit{Cohen's d}=0.082).  We conjecture this is because Kotlin/Java source files participating in the interactions are often committed to modify  Kotlin code related to the facade, such as method calls, field access, parameter changes.

We also identified 13 open-source projects containing at least 10 matched bug-commit records to analyze their cross-language characteristics. For each project, we calculated the average number of bugs and bug churn (modified lines of code to fix the bug) for files involved in cross-language dependencies against those that are not. The empirical results consistently demonstrate that files participating in cross-language interactions exhibit a significantly higher defect density and experience greater bug churn. By applying the SZZ approach~\cite{sliwerski2005szz,kim2006szz} to measure the lifespan of these defects, we found that in 10 of 13 projects, bugs involving files in CL interactions exist significantly longer before being resolved compared to non-CL bugs. This indicates that cross-language interactions not only tend to introduce more errors, but also make those errors more difficult to detect, leaving systems exposed to flaws for years.

\textbf{To answer RQ2, the average number of commits and churn for 40 open-source Kotlin-Java projects indicate that Kotlin source files participating in Kotlin-Java interactions tend to have more commits and more churns. Furthermore, these source files are significantly more bug-prone, suggesting that cross-language interactions increase maintenance complexity and make defects more likely to occur.} 

\subsection{Common Problematic Issues in Kotlin-Java Interactions}
\label{subsec:rq3}

RQ2's results reveal that Kotlin source files participating in Kotlin-Java interactions undergo more commits and are prone to changes. To answer RQ3, we delve into the details of why certain Kotlin-Java interactions are problematic. Following the approach described in Section~\ref{subsec:approach2}, we first located 2,915 commits involving certain code changes, which may signal problematic Kotlin-Java interactions. Then we manually inspected these commits and confirmed 308 commits containing fixes for 310 problematic Kotlin-Java interactions (note that in our previous conference paper~\cite{feng2024cross}, keyword searching identified 5,137 commits, of which 103 were true Kotlin–Java interaction patches, yielding a precision of 2.0\%. In this submission, using an improved \texttt{gumtree}-based method, we achieved a precision of 10.6\%). After discussion among three developers, we classified these 310 problematic Kotlin-Java interactions into 10 categories according to the issue and fixing strategies in Kotlin-Java problematic interactions. Table~\ref{tbl:smell} lists these categories and their occurrences. In the following, we provide an example for each of these 10 categories.


\begin{table}[!htbp]
    \centering
    \caption{Problematic Issues in Kotlin-Java Interactions}
    \label{tbl:smell}
    \begin{threeparttable}\begin{tabular}{lcc}
    \toprule
\textbf{Issue Category}  & \textbf{Occurrence}  \\ 
\midrule
Missing Nullability Annotation on Java      &  117    \\ 
Java Return Value NPE                               &  83  \\ 
Kotlin Inherit Java Nullability Inconsistency       &  38  \\  
Kotlin Wrong Cast                                    &  17  \\
Java Param Unknown Nullablity Passed into Kotlin  &  17  \\
Kotlin Wrong Not Null Assertion                       &  12  \\ 
Overuse Java Optional in Kotlin code                 &   7  \\ 
Immutable Collection Conversion                       &  4   \\
Missing Throws Annotations in Kotlin                  &  3   \\
Other & 12\\ 
\midrule
Total & 310 \\ 
\bottomrule
    \end{tabular}
    \end{threeparttable}
\end{table}

\subsubsection{\textbf{Missing Nullability Annotation on Java: fix by clarifying Nullability in Java code}}
\label{3-1}

This is the most common issue we observed among the 308 problematic interactions. When developing a Kotlin-Java system, adding appropriate nullable or non-nullable annotations (\texttt{@NotNull} or \texttt{@Nullable}) to Java's public API will enable modern IDEs to alert developers its nullability at the time of invocation and provide stricter code checking during compilation. Without annotations on parameters of Java's public API, Kotlin code is unable to determine the required nullability of parameters and an inappropriate API call will result in an NPE.


\begin{lstlisting}[
    style=bw-diff, 
    caption={Java API Add Annotation to Fix NPE},
    label={lst:rq3-1}
]
(*\textbf{+}~~~\textbf{@Nullable}\ *)
    public AudioDeviceInfo getCommunicationDevice() {
        return audioManager.getCommunicationDevice();
    }
\end{lstlisting}

Listing~\ref{lst:rq3-1} shows commit \href{https://github.com/signalapp/Signal-Android/commit/6c56ef470f0fa6889335675f1a52a240796ec9c4}{6ec9c4} from the {\sffamily Signal-Android} project, which is a good example of the importance of Java nullability annotation when interacting with Kotlin. To \dquote{\textit{Nullability safety for getCommunicationDevice().}} as stated in the commit comment, \texttt{@Nullable} annotation was added to the \textit{getCommunicationDevice} Java method and Kotlin code that invoked this method was changed accordingly.

For Kotlin developers who use Java libraries, due to the absence of nullability annotation in most Java libraries, we recommend that Kotlin developers carefully read the documentation and avoid passing nullable values into Java methods without confirming its nullability. We will discuss in Section~\ref{rq3:java-return} how to handle Java return values in Kotlin.

\subsubsection{\textbf{Java Return Value NPE (NullPointerException): fix by adding Null-safe Operator in Kotlin code}}
\label{rq3:java-return}

When Kotlin code calls a Java method and the Java method does not indicate the nullability of the return value with annotations such as \texttt{@NotNull} or \texttt{@Nullable} in the meantime, the Kotlin compiler will not be able to determine the nullability of the return value. If Kotlin code accesses a variable whose nullability is unknown, a NullPointerException (NPE) may get triggered.


\begin{lstlisting}[
    style=bw-diff, 
    firstnumber=75, 
    caption={Kotlin Add Null-Safe Operator to Fix NPE},
    label={lst:example5}
]
val Class<*>.shPackageName
    get() =
(*\textbf{-} \sout{~~~~~~~~        canonicalName.substringBeforeLast('.')}\ *)
(*\textbf{+} \textbf{~~~~~~~~        canonicalName?.substringBeforeLast('.')}\ *)
\end{lstlisting}

Listing~\ref{lst:example5} shows an example of this issue in commit \href{https://github.com/hannibal002/SkyHanni/commit/2cd4ca1}{2cd4ca1} in the {\sffamily SkyHanni} project. The crossed-out code attempts to call the method \texttt{getCanonicalName()} in the Java class `Class' and assumes that the returned value from this method is not \texttt{NULL}. However, at runtime this method returns a \texttt{NULL} and the crossed-out line throws an NPE. As stated in the commit comment, the commit aims to ``\textit{fix an NPE in ReflectionUtils.shPackageName}''.  It added the Null-Safe Operator \texttt{?} to the returned value of the \texttt{getCanonicalName()} method. In this way, the Null-Safe Operator \texttt{?} performs a \texttt{NULL} check on the returned value. If it is \texttt{NULL}, it stops the process without proceeding to the next call.

In this example, \texttt{getCanonicalName()} is code from the Java standard library that lacks Nullability Annotations, and its return type is treated by Kotlin as a platform type. Direct access on platform types is considered compile-time valid by the Kotlin compiler. Assuming \texttt{getCanonicalName()} were annotated with \textit{@Nullable}, \texttt{canonicalName.substringBeforeLast}  would be flagged as an error by the Kotlin compiler: \dquote{\textit{Direct access to members of nullable type is not allowed.}} As described in Section~\ref{3-1}, if the Java code used by Kotlin is maintained by the current developers, we only need to add Nullability Annotations and modify any Kotlin code that might cause compilation errors. However, in this example, since such Java code is not within the maintenance scope of the project, the Kotlin code can only be modified from direct access to null-safe access.

\subsubsection{\textbf{Kotlin Inherit Java Nullability Inconsistency: fix by ensuring parent-child API consistent in Kotlin code}}
\label{3-3}

This scenario happens when a Kotlin class inherits a Java class and overrides the parent methods, the nullability of parameters in the overridden methods is inconsistent between the parent and child class. This is not a problem in Kotlin-Kotlin or Java-Java interactions as the type consistency is manually enforced in modern IDEs. However, the nullability consistency is not enforced for Kotlin-Java interactions by modern IDEs so developers need to manually check it. The commit \href{https://github.com/nightscout/AndroidAPS/commit/d71cc391fef3a37405e465a0b2d88a71e09356cb}{9356cb} in the {\sffamily AndroidAPS} project shows an example and how developers were addressing it.

In Listing~\ref{lst:rq3-4-kotlin-inherit}, \texttt{ServiceTaskExecutor::} \texttt{afterExecute} method is written in Kotlin and it overrides the \texttt{ThreadPoolExecutor::afterExecute} Java method. The second parameter \texttt{t} in the Kotlin class \texttt{ServiceTaskExecutor} is not null while the parent Java class \texttt{ThreadPoolExecutor} does not specify the nullability of the second parameter which is actually \texttt{Nullable}. There exists an inconsistency between these two methods, when call the method \texttt{afterExecute} on an object with compile type of \texttt{ThreadPoolExecutor} but runtime type of \texttt{ServiceTaskExecutor}, passing a null value, an NPE occurs.


\begin{lstlisting}[
    style=bw-diff, 
    caption={Kotlin Inherit Java Nullability Inconsistency},
    label={lst:rq3-4-kotlin-inherit}
]
(*\textbf{-} \sout{~~~~override fun afterExecute(r: Runnable, t: Throwable) \{}\ *)
(*\textbf{+} \textbf{~~~~override fun afterExecute(r: Runnable, t: Throwable}\textbf{?}\textbf{) \{}\ *)
\end{lstlisting}

To fix this error, developers added the \textsc{null safe} operator \texttt{?} in the child Kotlin class \texttt{Throwable}, allowing the second parameter to be nullable. From this example, we can conclude that, although the inconsistency would not result in compilation errors within the child class itself, it can arise potential issues from a third class' calling. It is necessary to make sure that Kotlin-Java parent-child overridden methods strictly follow the same nullability type. Furthermore, this is also a violation of Liskov Substitution Principle, which requires a subtype should be substitutable for its supertype.

\subsubsection{\textbf{Kotlin Wrong Cast: fix by using safe cast or cast to nullable type in Kotlin code}}
\label{3-4}

This issue is common in the code of Java migration to Kotlin. Developers may confuse among these 3 type casting approaches: \ding{172} unsafe cast to Non-null type (\texttt{obj as Type}); \ding{173} safe cast(\texttt{obj as? Type} or \texttt{obj as? Type?}); \ding{174}  unsafe cast to Nullable type(\texttt{obj as Type?}). Notably, an unsafe cast to Non-null can result in an NPE when \texttt{obj} is \texttt{NULL} or result in a  \texttt{ClassCastException} (CCE) if \texttt{obj} is not the expected type \texttt{Type}. In contrast, the unsafe cast in Java, \texttt{(Type) obj} only raise a CCE, even when the object is \texttt{null}. Therefore, a semantically equivalent transformation during Java-to-Kotlin migration should convert \texttt{(Type) obj} to \texttt{obj as Type?}. The commit \href{https://github.com/nightscout/AndroidAPS/commit/34c4dc2a980fc62e0800ea4e7d5470ccf4c5ff69}{c5ff69} in the {\sffamily AndroidAPS} project shows an example and how developers were addressing it. 


\begin{lstlisting}[
    style=bw-diff, 
    caption={Kotlin Wrong Cast},
    label={lst:rq3-6-cast}
]
(*\textbf{-} \sout{(context?.getSystemService(NOTIFICATION\_SERVICE) as NotificationManager)\\.cancel(Constants.notificationID)}\ *)
(*\textbf{+} \textbf{(context?.getSystemService(NOTIFICATION\_SERVICE) as NotificationManager}\textbf{?}\textbf{)}\textbf{?}
    \\  \textbf{.cancel(Constants.notificationID)}\ *)
\end{lstlisting}

In Listing~\ref{lst:rq3-6-cast}, \texttt{context?.get...Service(...)} is a nullable value and unsafe cast to non-null type \texttt{NotificationManager} may lead to an NPE. To mitigate this risk, the developer resolved the issue by applying an unsafe cast to a nullable type instead. The commit message ``prevent NPE'' explicitly confirms that an NPE was encountered at this location. It's worth noting that since the migration from Java to Kotlin, this wrong cast issue has pervasively existed among many K-J cross-language projects (see more in commit \href{https://github.com/nightscout/AndroidAPS/commit/312d863e381891f65d9107f80b006f8b10a7704e}{a7704e}).

\subsubsection{\textbf{Java Unknown Nullablity Passed into Kotlin: fix by making Kotlin parameter nullable in Kotlin code}}
\label{3-5}

This issue is common when Java code is calling Kotlin methods. Most Java code has unknown nullablity, while Kotlin code must indicate nullability. When passing an unknown nullablity value into a Kotlin non-null parameter, the compiler or IDE will not give any error or warning. However, if the value is actually \texttt{NULL} in runtime, an NPE will be thrown by Kotlin. The commit \href{https://github.com/signalapp/Signal-Android/commit/be47e9e92869c1ba04db474cb70fda4be09e2c68}{9e2c68} in the {\sffamily Signal-Android} project shows an example and how developers were addressing it. 


\begin{lstlisting}[
    style=bw-diff, 
    caption={Make Kotlin Parameter Nullable},
    label={lst:rq3-8-kt-pub}
]
class IncomingMediaMessage(
(*\textbf{-} \sout{~~~~val from: RecipientId,}\ *)
(*\textbf{+} \textbf{~~~~val from: RecipientId}\textbf{?}\textbf{,}\ *)
      val groupId: GroupId? = null,
      val body: String? = null,
\end{lstlisting}

As shown in Listing~\ref{lst:rq3-8-kt-pub}, the non-null parameter \texttt{RecipientId} was changed to nullable in this commit. The commit message ``Fix NPE when receiving media only MMS'' proves that the developers faced an NPE here. 
In addition, we need to remind Java developers here that in Kotlin, parameter types without question marks (Non-null type) will immediately throw an NPE once they receive a \texttt{NULL} value. So Java developers should carefully check their arguments passing into a non-null Kotlin parameter.

\subsubsection{\textbf{Kotlin Wrong Not Null Assertion: fix by modifying non null assertion operator in Kotlin code}}
\label{3-6}

When a developer is confident that the returned value of a method or a property is non-null at all times, they apply a \textsc{not null} assertion \texttt{!!} operator in Kotlin code to emphasize its non-null nature. Such assertions offer significant convenience in subsequent code writing, eliminating the need for \texttt{NULL} checks and enhancing logical flow. However, non-null assertions also pose risks. If applied to a value whose nullability is not entirely determinable, an NPE may still occur at runtime if the value turns out to be null, potentially affecting a substantial portion of the subsequent code.


\begin{lstlisting}[
    style=bw-diff, 
    firstnumber=75, 
    caption={Kotlin Not Null Assertion Throws NPE},
    label={lst:example6}
]
val map: Map<ProtoMember, Any?>
(*\textbf{-} \sout{        get() = entries!!.toMap()}\ *)
(*\textbf{+} \textbf{        get() = entries?.toMap() ?: emptyMap()}\ *)
\end{lstlisting}

Listing~\ref{lst:example6} depicts commit \href{https://github.com/square/wire/commit/83dc35e}{83dc35e} from the {\sffamily wire} project. A \textsc{not null} assertion \texttt{!!} operator was placed on the \textit{entries} variable. However, this code throws an NPE at runtime because it fails to fully consider the nullability of the \textit{entries} variable. The commit message ``\textit{fix: options entries can be null for files imported from external jars}'' verified this observation and the \textsc{not null} assertion \texttt{!!} operator was changed to the null-safe Operator \texttt{?} in this commit.

\subsubsection{\textbf{Overuse Java \texttt{Optional} in Kotlin code: fix by leveraging safe check mechanisum in Kotlin code}}
\label{3-7}
One key feature of Kotlin is to eliminate the danger of null references by leveraging nullable types and non-nullable type. Compared to Kotlin, Java does not have such feature and thus it needs other techniques such as \texttt{Optional} to check whether a value is null or not. \texttt{Optional} is a container class introduced in Java 8 and provides a way to represent optional values without using null references, thereby helping to avoid common programming pitfalls associated with null values. The \texttt{Optional} class includes methods like \textit{isPresent()}, \textit{get()}, \textit{ifPresent()}, and \textit{orElse()}, which enable developers to check value presence and absence and to handle cases where a value might be missing.

We observed a case that Kotlin code overuse Java Optional, instead of leveraging Kotlin's own typing system. The commit \href{https://github.com/androidx/androidx/commit/0d6f2c284f31f4b61f0e5a8b6ac7970295f6a388}{5f6a388} in the {\sffamily androidx} demonstrates an example, with its commit comment ``\textit{Remove Optional from TypeSpec API}''.

Listing~\ref{lst:rq3-5-optional} illustrates a portion of the code modifications made to the \texttt{FieldBinding} Kotlin file. Previously, the \texttt{Optional} class was extensively utilized to encapsulate two properties: \textit{valueGetter} and \textit{valueSetter}. Overuse of the Optional class not only consumes additional memory and can negatively affect system performance, but it may also complicate code comprehension. We strongly suggest practitioners to examine the overuse of \texttt{Optional} in Kotlin code.


\begin{lstlisting}[
    style=bw-diff, 
    caption={Overuse Java Optional in Kotlin Code},
    label={lst:rq3-5-optional}
]
internal data class FieldBinding<T, BuilderT> 
    constructor(val name: String,
(*\textbf{-} \sout{~~~~val valueGetter: Function<T, Optional<Value>>,}*)
(*\textbf{-} \sout{~~~~val valueSetter: CheckedInterfaces.BiConsumer<BuilderT, Optional<Value>>}*)
(*\textbf{+} \textbf{~~~~val valueGetter: Function<T, Value}\textbf{?}\textbf{>,}*)
(*\textbf{+} \textbf{~~~~val valueSetter: CheckedInterfaces.BiConsumer<BuilderT, Value>}*)
) {
\end{lstlisting}
\subsubsection{\textbf{Immutable Collection Conversion: fix by creating a defensive copy of Java collections in Kotlin code}}
\label{3-8}

In Kotlin, collections are categorized into 2 classes: \ding{172} modifiable, e.g., \texttt{MutableList} or \texttt{MutableSet}; \ding{173} unmodifiable, e.g., \texttt{List} or \texttt{Set}. When Java code calls collections from Kotlin code, modern IDEs ensure that proper methods of collections are invoked, meaning that mutation methods like \textit{add()} or \textit{remove()} in Java cannot be applied to an unmodifiable Kotlin collection. 

However, if Kotlin code is accessing collections from Java code, the mutability of the collections is unknown to the Kotlin side. Since unmodifiable collections in Java function as a wrapper over existing modifiable collections, public methods such as \textit{add}, \textit{remove}, and \textit{update} in Java modifiable collections remain callable to Java unmodifiable collections. This presents a challenge on the Kotlin side, where the mutability of collections imported from Java cannot be determined statically.

Listing~\ref{lst:example4} illustrates an example of this issue in the commit \href{https://github.com/square/wire/commit/a8cf6eb}{a8cf6eb} of the {\sffamily wire} project. The \textit{getFromBuilder} method utilizes Java's reflection mechanism and returns an object with a java.util.Map type. Prior to this commit, this \texttt{Map} object was erroneously treated as a Kotlin \texttt{MutableMap}. Given that the mutability of Java's collections is not inherently recognized in Kotlin, developers introduced a check for this \texttt{Map} object in this commit. Records are added to it directly if the object is confirmed to be a \texttt{MutableMap}. On the contrary, once the object is confirmed to be a \texttt{Map}, a defensive copy of the map is created first, after which the records are put into the copied map. If the object failed to be cast to either a \texttt{MutableMap} or a \texttt{Map}, an exception is thrown. While this approach solves the immediate issue, using \texttt{obj is MutableMap} in JVM to check its mutability is not recommended. A more reliable strategy would be assuming that all collections of unknown mutability are immutable and create a defensive copy~\cite{youtrack-kt-39635}.


\begin{lstlisting}[
    style=bw-diff, 
    caption={Immutable Collection Conversion},
    label={lst:example4}
]
(*\textbf{-} \sout{val map = getFromBuilder(builder) as MutableMap<Any, Any>}\ *)
(*\textbf{-} \sout{map.putAll(value as Map<Any, Any>)}\ *)
(*\textbf{+} \textbf{when (val map = getFromBuilder(builder)) \{}\ *)
(*\textbf{+} \textbf{~~~~is MutableMap<*, *> -> map.putAll(value as Map<Nothing, Nothing>)}\ *)
(*\textbf{+} \textbf{~~~~is Map<*, *> -> \{}\ *)
(*\textbf{+} \textbf{~~~~~~~~val mutableMap = map.toMutableMap()}\ *)
(*\textbf{+} \textbf{~~~~~~~~mutableMap.putAll(value as Map<out Any?, Any?>)}\ *)
(*\textbf{+} \textbf{~~~~~~~~set(builder, mutableMap)}\ *)
(*\textbf{+} \textbf{~~~~\}}\ *)
(*\textbf{+} \textbf{~~~~else -> \{}\ *)
(*\textbf{+} \textbf{~~~~~~~~val type = map?.let \{ it::class.java \}}\ *)
(*\textbf{+} \textbf{~~~~~~~~throw ClassCastException("Expected a map type, got \$type.")}\ *)
(*\textbf{+} \textbf{~~~~\}}\ *)
(*\textbf{+} \textbf{\}}\ *)
\end{lstlisting}
\subsubsection{\textbf{Missing Throws Annotations in Kotlin: fix by adding \texttt{@Throws} Annotation in Kotlin code}} 
\label{3-9}

This situation arises because Kotlin does not support checked exceptions as defined in Java (see checked exception in Java Language Specification~\cite{java-spec}). The commit \href{https://github.com/signalapp/Signal-Android/commit/e5d196c64254b0cd55fa326d11a8e6ece684ab09}{84ab09} in the {\sffamily Signal-Android} project shows an example and how developers were addressing it. 

As we can see from Listing~\ref{lst:rq3-9-throws}, the Kotlin method \texttt{verifyFile} throws a Java checked exception \texttt{IOException}. But without the \texttt{@Throws} annotation, the Java compiler does not recognize that the method may throw a checked exception. Consequently, as not enforcing the standard compile-time checks for exception handling, runtime issues due to unhandled exceptions may occur.


\begin{lstlisting}[
    style=bw-diff, 
    caption={Missing Throws Annotations in Kotlin},
    label={lst:rq3-9-throws}
]
(*\textbf{+} \textbf{@Throws(IOException::class)}\ *)
  fun verifyFile(cipherStream: InputStream, 
    passphrase: String, expectedCount: Long): Boolean {
  ...}
\end{lstlisting}

\subsubsection{\textbf{Other}}
\label{3-10}

When the above nine categories exhibit clear patterns in issue characteristics and fixing strategies, changes in the \dquote{Other} category typically involve more complex fixes that cannot be addressed with just one or two simple modifications. In this category, developers must consider broader compatibility adjustments, such as those introduced by upstream changes that affect the entire project or by evolving requirements from downstream cross-language users. Also, the \dquote{Other} category is highly case-dependent, and the changes it contains are difficult to classify into the previously defined categories or to associate with any consistent pattern.




The commit \href{https://github.com/signalapp/Signal-Android/commit/53673be5cbd69c5664d5ab14c1f5b9aa14cecc32#diff-40e74dbbeb313a1c4036be9b93fc115e94150a363557e4ceab13dea36e94b92e}{94b92e} in {\sffamily Signal-Android} demonstrates such an issue developers encountered and the resolution process when upgrading from Kotlin 1.7.20 to Kotlin 1.8.10. Listing~\ref{lst:rq3-10-other-1} illustrates one line of code change in this commit. The method \texttt{Single.fromCallable} is written in Java and called from Kotlin. The curly braces represent the callback lambda function passed to \texttt{fromCallable}. The crossed-out line indicates that the return value of this lambda is \texttt{store.state.recipientId}. The type of this return value is a Java generic parameter whose generic argument is \texttt{RecipientId}. In Kotlin 1.7.20, when a Java generic parameter annotated with \texttt{@NotNull} has a non-null generic argument, it can accept values of nullable types. Here, although the generic argument is non-null \texttt{RecipientId}, the actual return type is inferred as a platform type \texttt{RecipientId}, meaning that the nullability of the return value is unknown. However, in Kotlin 1.8.10, the return type is strictly inferred as non-null \texttt{RecipientId}, which no longer allows returning \texttt{store.state.recipientId} whose nullability is nullable. Consequently, the developer added a non-null assertion (!!) to resolve this discrepancy.

Aside from this specific line of code, the commit includes changes to approximately 1,000 lines of source code, with many interacting files requiring diverse patterns of modifications. Therefore, we classify this commit with a Kotlin-Java interaction issue under the \dquote{Other} category.


\begin{lstlisting}[
    style=bw-diff, 
    caption={Upgrading from Kotlin 1.7.20 to 1.8.10 requires non-null assertion},
    label={lst:rq3-10-other-1}
]
return Single.fromCallable<RecipientId> {
        SignalStore.storyValues().
            userHasBeenNotifiedAboutStories = true
        Stories.onStorySettingsChanged(id)
(*\textbf{-} \sout{~~~~~~store.state.recipientId}\ *)
(*\textbf{+} \textbf{~~~~~~store.state.recipientId}\textbf{!!}\ *)
}.observeOn(AndroidSchedulers.mainThread())
\end{lstlisting}

\textbf{To answer RQ3, we conducted a manual inspection of 310 cases involving Kotlin-Java interactions, which had already led to system errors or incurred maintenance costs. Through these inspections, we identified 10 categories of common issues along with corresponding fixing strategies.} We believe that these categorized cases can serve as a taxonomy and assist developers in correctly handling Kotlin-Java interactions.
\subsection{Proactively Detecting Kotlin-Java Problematic Interaction Issues}
\label{subsec:rq4}

When Kotlin interacts with Java, a nullable variable from the Java side is converted to a platform type, which may then be further interpreted as a non-nullable type in Kotlin. This implicit conversion can lead to NullPointerExceptions (NPEs) at runtime, as discussed in Section~\ref{3-1} through Section~\ref{3-4}. \toolName{} leverages identified common patterns in these scenarios and implemented corresponding detection rules to proactively catch these issues.

As shown in Table~\ref{tbl:detect}, 248 out of 310 Kotlin-Java interaction issues were successfully detected. In particular, \toolName{} shows promise in detecting the first four, sixth, and seventh types of issues, which exhibit clear dependencies and identifiable code patterns. We manually inspected the undetected cases and found that the data flow in these scenarios is often complex, or that some cases may require dynamic analysis. Overall, \toolName{} demonstrated its effectiveness by detecting 80.0\% (248/310) of the Kotlin-Java interaction issues.

\begin{table*}[!htbp]
    \centering
    \caption{Detecting Results of Kotlin-Java Problematic Issues by \toolName{}}
    \label{tbl:detect}
    \begin{tabular}{lccc}
    \toprule
        \textbf{Issue Category} & \textbf{Occurrence} & \textbf{Detected} & \textbf{Reasons for Undetected} \\ \midrule
Missing Nullability Annotation on Java        &  117 & 117 & --- \\ 
Java Return Value NPE                         &  83  & 59  & The data flow related to nullability judgment is too complex. \\ 
Kotlin Inherit Java Nullability Inconsistency &  38  & 38 & --- \\  
Kotlin Wrong Cast                             &  17  & 17 & --- \\
Java Unknown Nullablity Passed into Kotlin    &  17  & 0 & The data flow related to nullability judgment is too complex. \\
Kotlin Wrong Not Null Assertion               &  12  & 10 & The data flow related to nullability judgment is too complex. \\ 
Overuse Java Optional in Kotlin code          &   7  & 7 & --- \\ 
Immutable Collection Conversion               &  4   & 0 & Cases here need dynamic analysis.\\
Missing Throws Annotations in Kotlin          &  3   & 0 & Cases here need dynamic analysis.\\
Other &12  & 0 & Cases too complex. \\ 
\midrule
Total &310 &  248 & --- \\ 
        \bottomrule
    \end{tabular}
\end{table*}

\textbf{To answer RQ4, \toolName{} takes source code as input and successfully detects 248 out of 310 Kotlin-Java interaction issues, demonstrating its effectiveness.}
\section{Discussions}
\label{sec:discussion}

\subsection{Analysis of Results}
The finding in \textbf{RQ1} - that dependencies between Kotlin and Java are prevalent in Kotlin–Java projects - is supported by the fact that the two languages are designed to interoperate directly at the source-code level; for example, Kotlin and Java can freely \texttt{access} each other’s fields, methods, and classes. In contrast, Python and C++ typically interact only through explicitly exposed APIs, so their cross-language interactions occur at a much narrower interface boundary. The widespread, fine-grained interactions between Kotlin and Java require developers to understand the features of both languages, introducing challenges that go beyond simply calling API-level functions. The results of \textbf{RQ2} indicate that source files involved in Kotlin-Java interactions undergo more commits, have higher defect density, greater maintenance churn, and significantly longer bug lifespans, than those exclusively involved in a single language. It demonstrates that Kotlin-Java interfaces are inherently more active, just like other cross-language interfaces, even though Kotlin and Java can coexist within the same IDEs and their interaction mechanism is different from other cross-language systems.

The manual inspections of 310 cases in \textbf{RQ3} reveal specific 10 categories of issues within Kotlin-Java interactions. Upon further classification, we found that 284 out of 310 problematic interactions between Kotlin and Java are related to NullPointerExceptions (NPEs). Despite Kotlin being known for its better null safety mechanism compared to Java, the interface between Kotlin and Java still experiences NPEs. Most Kotlin/Java interaction problems are related to NPEs - aligns with the fact that differences between the Kotlin and Java type systems are the root cause of many nullability issues in cross-language contexts. By contrast, Python/C++ interfaces tend to encounter more vulnerability-related problems~\cite{li:2022polycruise}. For example, Python is memory-safe while C++ is not, and their API boundary often exposes risky manual memory operations. In comparison, Kotlin and Java do not face such memory-safety concerns because both rely on the JVM’s garbage collection mechanism. Other issues, such as ``Immutable Collection Conversion'' and ``Overuse of Java Optional in Kotlin'', result from unfamiliarity with Kotlin documentation. It is also surprising that we only find 3 exception handling issues in the Kotlin-Java interactions, despite Java and Kotlin having different exception catching rules. This can be attributed to modern IDEs' checking in the compilation process. 

Furthermore, the results of \textbf{RQ4} show that the misalignment between the Kotlin and Java type systems is the root cause of most NPE issues in Kotlin-Java interactions. With well-designed enforcement rules, static analysis of source code can effectively detect these issues. This also suggests that modern IDEs and Kotlin/Java compilers could incorporate such static checks into their systems to achieve a better interoperability. As our next step, we also plan to improve the accuracy and scalability of the tool and explore integrating the tool into CI pipelines or IDE plugins, making it more suitable for automated usage in practice.

\subsection{Implications}
\textbf{To Software Practitioner and IDE developers:} Although Kotlin and Java can interact smoothly at the code level, common interaction issues still exist that may not be captured by modern IDEs. The identified categories of interaction issues and their fixing strategies can serve as a valuable reference, complementing the Kotlin-Java interoperability guide from Android~\cite{android-kotlin-java-interop}. IDE developers can also consider to incorporate more static Kotlin-Java checks to warn the Kotlin-Java interaction risk. Furthermore, based on our observations of Kotlin–Java problematic interaction issues, we submitted a PR (see the \href{https://github.com/JetBrains/kotlin-web-site/pull/4754\#issuecomment-3351136668}{link}) to the official Kotlin documentation, adding a warning about casting Java types to Kotlin, which can help developers migrating from Java to Kotlin avoid such errors. Although the PR was not merged in its initial form due to a need for more comprehensive context and refinements discussed during the code review thread, the core findings were validated by the official Kotlin team. Consequently, they explicitly incorporated partial documentation updates regarding safe type casting into the official Kotlin documentation guide (https://kotlinlang.org/docs/java-to-kotlin-nullability-guide.html\#casting-types-safely) based on our findings.

Also, the fact that in most projects cross-language (CL) bugs evade detection for years proves they cannot be caught by standard compilers or localized testing. Developers should not rely on standard unit tests; instead, they must implement strict interface contracts, write dedicated cross-language integration tests, and mandate rigorous peer reviews for all multi-language pull requests to prevent these long-lived defects. Our \toolName{} can assist in detecting these issues during this process.

\textbf{To Researchers:} Different cross-language interactions may suffer from various problems. Conducting empirical studies to identify problematic interactions with associated errors and maintenance costs is key to pinpointing them accurately. For Kotlin-Java interactions, there is a high demand for research and tools that can automatically detect and fix these issues.

\subsection{Threats to Validity}
The first threat is the accuracy of our dependency extraction tool. \toolDepend{} is based on the Kotlin compiler and dependency types are defined according to previous research. Though we built the ground truth of a Kotlin-Java project and validated our tool can achieve a relatively high precision and recall, we cannot guarantee that it is 100\% accurate. We will continue to validate and improve our tool's accuracy in the future.

A key threat stems from the false positives generated by \toolName{}, which are driven by two main factors: language-level type system mismatches (where unannotated Java variables default to ambiguous platform types) and the tool's current reliance on surface-level semantic analysis rather than deep inter-procedural data-flow tracing. Consequently, establishing absolute ground truth currently requires significant manual inspection of natural-language Javadocs and project-specific contexts, introducing potential human verification overhead. To mitigate this threat, \toolName{} currently operates strictly as an advisory static linter while we actively discuss compiler plugin integration with the JetBrains team. Furthermore, incorporating an LLM-based documentation scraper to automatically infer platform-type nullability and a more sophisticated data-flow analysis engine to track object lifecycles across cross-language interface remains part of our future work.

Another threat is that we associated \#commit and Churn with file paths. A file involved in K-J dependencies may undergo activities unrelated to fixing interaction problems, or these activities may be mixed in a commit. We acknowledge that \#commit and Churn may not capture the exact maintenance cost of a file. However, there is no practical way to split the cost and directly measure the exact amount. We hope that the qualitative analysis detailed in Section~\ref{subsec:rq3} can partly mitigate this threat.


An external validity is that we only investigated 40 real-world projects with Kotlin as the primary languages. It is possible that additional categories of problematic Kotlin-Java or Java-Kotlin interactions may be discovered in other projects. Our future work will aim to include more software systems and industry examples to further validate our findings.
\section{Related Work}
\label{sec:related}

In this section, we discuss related research to our work from two perspectives: empirical studies on Kotlin-Java projects and code dependency extraction in cross-language systems.

\subsection{Empirical Studies on Kotlin-Java Projects}
Since the introduction of Kotlin, various empirical studies have been conducted to understand Kotlin-Java projects. Martinez and Mateus investigated why developers migrate Android applications from Java to Kotlin~\cite{martinez:2021tse}. They also examined the adoption, usage, and evolution of Kotlin features in Android development~\cite{mateus:2020esem}. Nakamura \textit{et al}. conducted a performance study of Kotlin and Java programs using bytecode analysis~\cite{nakamura:2024performance}. Ardito \textit{et al}. compared the effectiveness of Kotlin versus Java in Android app development tasks~\cite{ardito:2020ist}. Flauzino \textit{et al}. compared code smells between Java and Kotlin~\cite{flauzino:2018you}. Bose \textit{et al}. compared Java versus Kotlin in terms of null safety, exception handling, and other aspects~\cite{bose2018:comparative}.

While previous studies have compared various characteristics of Java with Kotlin, our study offers an empirical investigation into how Kotlin actually interacts with Java code. Despite the availability of documentation from Google and Android guiding Kotlin-Java interaction, to our knowledge, our study is the first to examine Kotlin-Java interactions in real-world projects. Our findings reveal that while Kotlin and Java demonstrate great interoperability and interact with each other extensively across systems, challenges and issues persist in such interactions. The recurring patterns identified in problematic Kotlin-Java interactions in this paper deserve special attention from developers, especially Kotlin developers.

\subsection{Code Dependencies Extraction}
The dependency relations among entities in the source code form the foundations of software  analysis~\cite{mo:2019tse,cui:2019icse,xiao:2021tse}. Existing techniques often rely on static code analysis to capture and model extract dependency relations from source code such as Java, Python and C/C++ ~\cite{jin:2024pyanalyzer,filgueira:2022inspect4py,molina:2021automatic, savic:2014language}. Various tools such as Structure 101~\cite{structure101}, Understand~\cite{understand}, and DV8~\cite{dv8} have been developed to support the extraction of dependency relations among entities in large-scale software projects.

While techniques and tools for extracting dependencies among code entities within single programming language are relatively mature, the extraction of dependencies among multiple programming languages remains largely unexplored. With advancements in modern platforms, it has become common to utilize multiple programming languages within the same system to achieve enhanced performance and scalability~\cite{mayer:2017multi,tomassetti:2014empirical}. However, this also increases the complexity of systems, posing challenges for extracting dependencies in such heterogeneous langauge environments. A recent systematic review reveals that traditional code analysis tools face challenges in these cross-language systems~\cite{latif:2023pragmatic}. A study also highlights high-risk vulnerabilities between Python and C interfaces, proposing symbolic dependence analysis to detect such vulnerabilities~\cite{li:2022polycruise}. As a complement to the above studies, our study proposed to extract code dependencies between Java and Kotlin code and further detect problems in Kotlin and Java interactions.

\section{Conclusions}
\label{sec:conclusion}

In this paper, we conducted an empirical study of Kotlin-Java interactions in 40 real-world projects. Our results demonstrate that Kotlin and Java interact frequently, confirming the extensive interoperability of Kotlin and Java. However, our study also reveals that source files involved in such cross-language systems have more commits and more defects compared to those involved in only a single language.

Furthermore, we identified nine common problematic scenarios in Kotlin-Java interactions, along with their corresponding fixing strategies. These problematic issues were extracted from commits in Git repositories and have been shown to require more maintenance effort. We further implemented a prototype called \toolName{} to proactively detect these problematic scenarios from source code.  We believe that recognizing and detecting these identified issues in Kotlin-Java interactions can assist developers in effectively improving such cross-language interactions. 

Our future work will focus on the following: first, investigating additional industry examples to further validate our findings; second, improving and integrating our \toolName{} tool with modern IDEs such as IntelliJ IDEA (e.g., as a plugin or through a proposal to the IDEA team) to help prevent Kotlin-Java interaction issues; and finally, enhancing our \toolDepend{} tool to support a broader range of cross-language dependencies.

\section*{Acknowledgments}
This work has been partially sponsored by the National Natural Science Foundation of China (NSFC) under Grant No. 92582203. The numerical calculations in this paper have been done on the supercomputing system in the Supercomputing Center of Wuhan University.

\balance
\bibliographystyle{IEEEtran}
\bibliography{reference}

\end{document}